\documentclass[12pt]{iopart}
\usepackage{iopams}  

\usepackage{graphicx}
\usepackage{setstack}
\usepackage{appendix}
\usepackage{enumerate}
\usepackage{epsfig}
\usepackage{times}
\usepackage{verbatim}
\usepackage{bbm}
\usepackage{color}

\definecolor{jens}{rgb}{0,.8,.5}
\newcommand{\je}[1]{{\color{jens} #1}}

\newcommand{\be}{\begin{equation}}
\newcommand{\ee}{\end{equation}}
\newcommand{\bea}{\begin{eqnarray}}
\newcommand{\eea}{\end{eqnarray}}
\newcommand{\ba}{\begin{align}}
\newcommand{\ea}{\end{align}}

\newcommand{\1}{\mathbbm{1}}

\newcommand{\avr}[1]{\left \langle#1 \right \rangle}

\newcommand{\bR}{\mathbbm{R}}

\newcommand{\cH}{\mathcal{H}}
\newcommand{\cD}{\mathcal{D}}

\newcommand{\cL}{\mathcal{L}}

\newcommand{\cZ}{\mathcal Z}
\newcommand{\cO}{\mathcal O}
\newcommand{\cQ}{\mathcal Q}

\newcommand{\half}{\frac{1}{2}}

\newcommand{\Ent}{{\rm Ent}}
\newcommand{\Var}{{\rm Var}}
\newcommand{\Cov}{{\rm Cov}}

\newtheorem{theorem}{Theorem}
\newtheorem{lemma}[theorem]{Lemma}
\newtheorem{corollary}[theorem]{Corollary}
\newtheorem{proposition}[theorem]{Proposition}
\newtheorem{definition}[theorem]{Definition}

\def\Proof{\noindent\textsc{Proof:}}
\def\proof{\Proof}
\def\qed{\leavevmode\unskip\penalty9999 \hbox{}\nobreak\hfill
     \quad\hbox{\leavevmode  \hbox to.77778em{%
               \hfil\vrule   \vbox to.675em%
               {\hrule width.6em\vfil\hrule}\vrule\hfil}}
     \par\vskip3pt}
    {\hspace*{\fill}$\Box$\vspace{1.5ex}\par}

\begin{document}

\title{Rapid mixing implies exponential decay of correlations}
\date{\today}
\author{Michael J. Kastoryano and Jens Eisert}
\address{QMIO Group, Dahlem Center for Complex Quantum Systems, Freie Universit{\"a}t Berlin, 14195 Berlin, Germany}
\vspace{-0.0cm}
\ead{mkastoryano@gmail.com}
\vspace{-0.2cm}

\begin{abstract}
We provide an analysis of the correlation properties of spin and fermionic
systems on a lattice evolving according to open system dynamics generated by a local primitive Liouvillian. We show that if the Liouvillian has a spectral gap which is independent of the system size, then the correlations between local observables decay exponentially as a function of the distance between their supports.  We prove, 
furthermore, that if the Log-Sobolev constant is independent of the system size, then the system satisfies clustering of correlations in the mutual information -- 
a much more stringent form of correlation decay.  As a consequence, in the latter case we get an area law (with logarithmic corrections) for 
the mutual information. As a further corollary, we obtain a stability theorem for local distant perturbations.
We also demonstrate that gapped free-fermionic systems exhibit clustering of correlations in the covariance and in the mutual information.  
We conclude with a discussion of the implications of these results for the classical simulation of open quantum systems with matrix-product operators and 
the robust dissipative preparation of topologically ordered states of lattice spin systems. 
\end{abstract}
\vspace{-1.2cm}



\section{Introduction}

Recent years have seen enormous progress at the interface between quantum information theory, condensed matter, and statistical physics. The fundamental formulation of many of our physical theories describes very completely the behavior of single, or small collections of, objects. But when many particles act together, there are a number of emergent phenomena which are not encoded in the fundamental constituents or dynamics in any manifest way. Important examples of these phenomena are: finite speed of propagation of information, irreversibility, or locality of correlations.
Lately, a number of remarkable results have been obtained which rigorously analyze the consequences of the locality of interactions of quantum many-body systems described by a Hamiltonian reflecting finite-ranged interactions. 
For a long time it has been common folklore  in the condensed matter community that gapped\footnote[1]{The absolute difference between the smallest and second smallest eigenvalues of the Hamiltonian remains fixed as the system size grows.} many body systems  do not have long range correlations (delocalized excitations). 
Recently this intuition has been made precise in a number of different ways for finite dimensional quantum systems. In particular, it has been shown that for a fixed dimension,  the ground state of a gapped local Hamiltonian exhibits clustering of correlations \cite{HastingsCC,NachtergaeleCC}.
An alternative notion of locality of correlations is related to the amount of correlation that a subsystem shares with its complement. If these correlations are proportional to the boundary, then the state is said to satisfy an \textit{area law}. 
It has been shown that one dimensional gapped systems, with a not overly degenerate ground state subspace, satisfy an area law \cite{AreaReview,AreaLaw1,AreaLaw3} for the entanglement entropy in the ground state. In turn, this often allows for an efficient classical description of these quantum states 
in terms of matrix-product states \cite{AreaLaw1,MPS}. It turns out also, that in one dimension, clustering of correlations alone already implies an area law for pure states \cite{BrandaoAreaLaw}. Rigorous results in two or more dimensions are still very scarce, with results for free fermionic and
bosonic models being an exception \cite{AreaReview,OurCMP}. Importantly, it has also been shown that local gapped (frustration-free, local topologically ordered) Hamiltonians are stable to local perturbations \cite{Mikalakis1,Mikalakis2,Sims}. Very recently, these stability results have been extended to the open system setting under appropriate conditions \cite{cubitt}.

In this work, we extend some of the results mentioned previously to a specific class of open quantum systems (primitive semi-groups), which include as a special case thermal quantum semi-groups (Davies maps \cite{Davies,Spohn}, see Appendix A). Our work may also be seen as a quantum generalization of a series of  results which analyze the connection between dynamical and static properties of classical spin systems evolving in time via Glauber dynamics (Metropolis or heat bath) \cite{Martinelli1,CombinMixing, ZiggyGuionet,Yoshida}.

In particular, we show how global mixing properties of the semi-group impose strong restrictions on the types of correlations which can be found in the steady state.  The functional methods developed in Refs.~\cite{LogSobolev,chi2} distinguish between two fundamental convergence behaviors: an asymptotic convergence rate which decreases with the system size and goes to zero in the thermodynamic  limit (non gapped), and an asymptotic rate which is lower bounded for all system sizes (gapped). We will focus on the second case, and point out that there are two subclasses in the analysis. The first  (Log-Sobolev constant) allows us to certify that the dissipative system converges in a time of order $\log N$, where $N$ is the number of sites in the system, whereas the second one ($\chi^2$ constant) only allows one to certify that the mixing time is of order $N$. 
The main result of this work is  a strong connection between the mixing behavior of rapidly mixing semi-groups of lattice systems,  and the correlations in their steady state. We show that if the semi-group is mixing in the strongest sense ($\log N$), then the correlations in the stationary state, as measured by the mutual information, decay exponentially in the distance between observables, whereas if the semi-group mixes in a time ($N$), then the correlations decay exponentially in the covariance. Mutual information correlation decay is shown to be much stronger that covariance correlation decay for spin systems. In particular, this allows us to establish and area law (with logarithmic corrections) in the mutual information. For free-fermionic systems, there is only one time scale of rapid mixing, in that if the system has a constant gap, then the correlations decay exponentially in the mutual information. Finally, for the proof of clustering of correlations in the mutual information for spin systems, we show a stability result for the stationary state under local Liouvillian perturbations which corroborate the recent results in Ref.~\cite{cubitt}. Our results can be seen as a generalization and completion of the sketch presented in Ref.~\cite{LRbound2}, which first discussed the question of the clustering of correlations due to dissipative Markovian dynamics. In particular, we provide rigorous tight bounds for the clustering of correlations of stationary states in the presence of a Liouvillian gap.

It is worth noting that in the classical setting, a stronger correspondence has been shown; namely that clustering of correlations implies a Log-Sobolev constant, for suitable boundary conditions \cite{Martinelli1,CombinMixing, ZiggyGuionet, Yoshida}. This essentially shows that, within the Glauber dynamics setting, the notions of rapid mixing and short range correlations are essentially equivalent.

This work is organized as follows: In section \ref{sec:setting}, we set the notation for lattice spin systems and for free fermionic models; we also define properties of Liouvillians; in section \ref{sec:mixing} we review the basic mixing properties of primitive semi-groups; 
in section \ref{sec:CC}, we define and compare different notions of clustering of correlations in the open system setting; in section \ref{sec:main} we state and prove the main results of 
this work which prove the relationship between rapid mixing and clustering of correlations; and finally in section \ref{sec:concl} we conclude and provide an outlook by discussing the implications of our result for the robust preparation of topologically ordered states in lattice systems. 
Throughout, we we will try to keep the formalism as elementary as possible, focussing on the physical motivations and  main mechanisms underlying the proofs. We will however try to point out whenever generalizations are possible.

\section{Formal setting}\label{sec:setting}

Throughout, we will consider finite dimensional quantum systems arranged on a lattice. For clarity of presentation we will consider a square lattice in $\cD\leq 3$ dimensions, but the results presented here can easily be extended to more general lattice systems. Let $\Lambda$ be the collection of lattice sites, then for any $z\in\Lambda$, let $\cH_z$ be the local Hilbert space associated with site $z$. For some subset $A\subset\Lambda$ of the lattice, we define the subsystem Hilbert space as 
\be 
	\cH_A=\bigotimes_{z\in A} \cH_z.
\ee
The dimension of the entire Hilbert space will be denoted $d$, while the dimension of the Hilbert space reduced to subsystem with support on $A$ will be written $d_A$. 
States are given by density matrices $\rho$, and reduced states are given by the partial trace restricted to a certain subsystem: 
$\rho_A=\tr_{A^c}({\rho})$, 
where $A^c$ is the complement of $A$. Observables are Hermitian operators $f=f^\dag$ denoted by lower case roman letters.  In a slight abuse of notation and language, we will refer to both the subset of lattice sites $A\subset\Lambda$ and to the space spanned by the local Hilbert spaces defined on those sited by $A$. We also will say that an operator $f$ is supported on $A$, to mean that $f$ is supported on the Hilbert space spanned by the lattice sites associated with $A$. 

The (dissipative) dynamics which we consider are described by a quantum dynamical semi-group, which often accurately approximates weak system environment couplings in the Markovian limit; as is the case in many quantum optics setups. The \textit{Liouvillian} (generator) of the semi-group is given by $\cL(\rho):=\dot{\rho}=i[H,\rho]+\cD(\rho)$, where
\be 
\cD(\rho)=\sum_j \left(L_j \rho L^\dag_j-\half(L^\dag_j L_j\rho+\rho L_j^\dag L_j)\right).
\ee
Time evolved states will be written $\rho_t=e^{t\cL}(\rho)$ and similarly for observables $f_t=e^{t\cL^*}(f)$, where $\cL^*$ is the Liouvillian in the Heisenberg picture.

We will say that a Liouvillian has support on a subset $V\subset \Lambda$ of the lattice, if $\cL$ acts trivially on all elements outside of $V$. $\cL$ is said to be local if it can be written as a sum of terms with local support: i.e., 
\begin{equation}
\dot{\rho}=\sum_{A\subset\Lambda}
\left(i[H_A,\rho]+\cD_A(\rho)\right), 
\end{equation}
where each $\cD_A$ has support on a subset $A\subset\Lambda$ independent of the system size (and typically small). 
Such local Liouvillians have been in the focus of intense recent research \cite{LRbound2,Diehl,Cirac,Prosen1,Prosen2,Diehl2,MPO3,Kliesch,LRbound1,Gadgets},
both with respect to condensed-matter inspired questions and applications in quantum information science.

Throughout this work, we will restrict ourselves to Liouvillians with norm-bounded local interactions. More specifically, we assume that $\cL$ is a sum of local terms with norm bound $L:= \sup_{Z\subset\Lambda}\|\cL_Z\|$, maximum range $a:=\sup_{Z:\cL_Z\neq0} {\rm diam}(Z)$, and maximum number 
\begin{equation}
	\cZ:=\max_{Z:\cL_Z\neq0}|\{ Z'\subset\Lambda|\cL_{Z'}\neq0,Z'\cap Z\neq \emptyset\}|
\end{equation}	
of nearest neighbors,
where ${\rm diam}(Z):=\max_{x,y\in Z}d(x,y)$ is the diameter of $Z$ and $d(\cdot,\cdot)$ is a metric on the lattice $\Lambda$ 
(typically the manhattan metric), and $\|\cdot\|$ is the operator norm. A Liouvillians is said to be \textit{primitive} \cite{wielandt}, if it has a unique full 
rank stationary state. Thermal Liouvillians (see Appendix A) and generic Markovian noise are examples of primitive Liouvillians. 
A primitive Liouvillian $\cL$ will be called \textit{regular} if for any bipartite subset $AB\in\Lambda$, the Liouvillian $\tilde{\cL}=\cL-\sum_{Z: Z\cap\partial_{AB}\neq0}\cL_Z$ obtained by removing all of the terms intersecting the boundary $\partial_{AB}$ is also primitive. Although this assumption might seem odd, it turns out to be very natural when discussing scaling of a dissipative system with the system size. In particular, primitive translationally invariant systems satisfy this property. 

We also introduce the notion of reversibility \cite{DB} (i.e.,  detailed balance) for primitive Liouvillians. To define reversibility, we will need to introduce a family of multiplication operators. Let $\sigma>0$ be a full-rank density matrix, then we define 
\begin{equation}
	\Gamma^s_\sigma(f)=\half(\sigma^s f \sigma^{1-s}+\sigma^{1-s} f \sigma^s) 
\end{equation}
for any observable $f$, and $s\in [0,1]$.  We say that a Liouvillian is 
\textit{s-reversible} (with respect to $\sigma$) if
\be  \Gamma^s_\sigma \cL^* = \cL \Gamma^s_\sigma.\label{detailed balance}\ee 
Note that Eq.\ (\ref{detailed balance}) is equivalent to $\avr{f,\cL^*(g)}_{s,\sigma}=\avr{\cL^*(f),g}_{s,\sigma}$, with the inner product 
\begin{equation}\label{spd}
	\avr{f,g}_{s,\sigma}=\tr[\Gamma^s_\sigma(f^\dag)g]. 
\end{equation}
Hence, if $\cL$ is s-reversible for some $s\in[0,1]$, then its spectrum is  real.  
 It is worth mentioning that {\it thermal} (as well as {\it Metropolis} \cite{Metropolis}) Liouvillians are s-reversible for all $s\in[0,1]$, as can easily be derived from the definition of reversibility and Eqs.\ (\ref{DBDavies1}) and (\ref{DBDavies1}) in the Appendix.

We will specifically mention also \textit{free-fermionic Liouvillians}, as they prominently feature in recent
studies of noise-driven criticality and
topological order by dissipation \cite{Prosen1,Prosen2,Diehl,Diehl2}. 
Such systems are described by $2N$  \textit{Majorana fermions} $r_1,\dots, r_{2N}$ for $N$ modes, one associated with each lattice site, 
where $N=|\Lambda|$. These Majorana operators
satisfy the anti-commutation relations $\{r_j,r_k\} = \delta_{j,k}$, $j,k=1,\dots, 2N$, and  
can be collected in a row vector $r=(r_1,\dots, r_{2n})^T$. In free fermionic open quantum systems the Hamiltonian
\begin{equation}
	H= i r^T h r
\end{equation}
with $h=- h^T\in \bR^{2n\times 2n}$
is taken to be a quadratic polynomial in these Majorana fermions, 
while each of the $L_j$ is a linear polynomial. The Liouvillian $\cL$ is local if both $H$ is local (in the sense that it is a sum of terms supported on geometrically local
modes) and each of the $L_j= l_l^T r$ 
is supported on a small number of sites only. The stationary states of such Liouvillians are Gaussian fermionic states, entirely captured in terms of 
the covariance matrix $\gamma\in \bR^{2N\times 2N}$ having entries $\gamma_{j,k} = i \tr(\rho {[}r_j, r_k{]})$. In the same way, covariance matrices of 
subsystems can be defined. The covariance matrix satisfies 
$\gamma = -\gamma^T$ and $-\gamma^2\leq \1$ \cite{Prosen1,Prosen2,Diehl2,OurCMP}.

\section{Mixing times of semi-groups}\label{sec:mixing}

In this section, we review tools from the theory of mixing times of primitive semi-groups, which will be necessary for the main results of this work. For a more comprehensive exposition, consult Ref.~\cite{LogSobolev}.
The mixing time of a quantum Markov process is the time it takes for the process to become close to the stationary state, starting from an arbitrary initial state. A huge amount of effort has been invested in bounding the mixing time of classical Markov processes; especially in the setting of Markov chain Monte Carlo \cite{MCMC}. 
Recently, a set of functional tools have been developed for analyzing the quantum analogue of Markov chain mixing for one parameter semi-groups \cite{chi2,LogSobolev}. In particular, trace norm convergence of primitive semi-groups can be very well characterized in terms of two quantities: the inverse of the smallest eigenvalue of the stationary state 
$\|\sigma^{-1}\|$, and one of two exponential decay
rates, the $\chi^2$  constant $\lambda_s$ \cite{chi2} or the Log-Sobolev constant $\alpha_s$ \cite{LogSobolev}. Each of these quantities has a convenient variational characterization. Let $\cL$ be a primitive Liouvillian, then for any choice of $s\in[0,1]$ define
\bea
 \lambda_s &:=& \inf_{f=f^\dag} \left. -\frac{d}{dt}\log[\Var^s_\sigma(f_t)]\right|_{t=0},\label{variance}\\
\alpha_s &:=& \inf_{f>0} \left.- \frac{d}{dt}\log[\Ent^s_\sigma(f_t)]\right|_{t=0}\label{entropy},
\eea where 
\bea\Var^s_\sigma(f)&=&|\tr[f\Gamma^s_\sigma(f)]-\tr[\sigma f]^2|,\\
\Ent^s_\sigma(f)&=& \tr[\Gamma^s_\sigma(f)(\log(\Gamma^s_\sigma(f))-\log(\sigma))] -\tr[\Gamma^s_\sigma(f)]\log(\tr[\Gamma^s_\sigma(f)]),
\eea
are variance and entropy functionals, and $f_t=e^{t\cL^*}(f)$ is the time evolved operator $f$ with respect to the Liouvillian $\cL$. Eqs.\  (\ref{variance}) and (\ref{entropy}) are generalizations of similar expressions defined in Ref. \cite{LogSobolev}, where only the $s=1/2$ case was considered. Given that $\Gamma^s_\sigma$ is only (completely)-positive for $s=1/2$, certain results only hold for that case. Manipulation of the covariance functional is most convenient for $s=0$. When $\cL$ is $s$-reversible for some $s\in [0,1]$, its  $\chi^2$ constant is equal to the spectral gap\footnote{The largest non-zero real part of an eigenvalue of $\cL$. Note that the real part of $\cL$ only takes non-positive values.} of $\cL$ -- which is usually associated to the convergence rate of a process. Indeed, 
the following is true.

\begin{lemma}[Spectral gap]\label{sLem}
Let $\cL$ be a primitive Liouvillian with stationary state $\sigma$, and suppose that it is s-reversible for some $s\in[0,1]$. Then $\lambda_s=\lambda$, where $\lambda$ is the spectral gap of $\cL$.
\end{lemma}
\proof{
Given that $\cL$ is s-reversible, $\Gamma^s_\sigma \cL^*= \cL \Gamma^s_\sigma$. Equivalently, 
\be 
	(\Gamma^s_\sigma)^{1/2} \cL^* (\Gamma^s_\sigma)^{-1/2}= (\Gamma^s_\sigma)^{-1/2}\cL (\Gamma^s_\sigma)^{1/2}=: \hat{\cL}_s,
\ee 
where $\hat{\cL}_s$ is clearly a hermicity preserving super-operator, given that $\Gamma_\sigma^s$ is hermiticity preserving. 
Because $\hat{\cL}$ and $\cL$ are related by a super-operator similarity transformation, they both have the same spectrum, and in particular the same spectral gap. 
Noting that the fixed point of $\hat \cL_s$ is $\sqrt{\sigma}$, the gap of $\cL$ can be written in terms of its variational characterization as
\bea 
	\lambda &=&\inf_{f=f^\dag, \tr[f\sqrt{\sigma}]=0} \frac{-\avr{f,\hat{\cL}_s^*(f)}}{\avr{f,f}}\label{gap1}.
\eea	
Given that $(\Gamma^s_\sigma)^{1/2}$ is bijective and hermicity preserving, for each such $f$ there exists a hermitian $g$ such that
\bea 
	f= (\Gamma^s_\sigma)^{1/2}(g).
\eea

Therefore, we can formulate the gap as
\bea 
	\lambda &=&\inf_{g=g^\dag, \tr[g \sigma]=0} \frac{-\avr{(\Gamma^s_\sigma)^{1/2}(g),\hat{\cL}_s^*((\Gamma^s_\sigma)^{1/2}(g))}}
	{\avr{(\Gamma^s_\sigma)^{1/2}(g),(\Gamma^s_\sigma)^{1/2}(g)}}.
\eea
That is to say, using the scalar product of Eq.\ (\ref{spd}),
\bea 
	\lambda &=&\inf_{g=g^\dag, \tr[g \sigma]=0} \frac{-\avr{g,\cL_s^*( g)}_{s,\sigma}  }
	{\avr{g,g}_{s,\sigma}} \\
	&=& \inf_{g=g^\dag}  \frac{-\avr{\tilde g,\cL_s^*(\tilde g)}_{s,\sigma}  }
	{\avr{ \tilde g,\tilde g}_{s,\sigma}} ,
\eea
defining $\tilde g = g - \tr[g\sigma]$, in such a way that $\tr[\tilde g  \sigma]=0$ for all $g$. We now make use of the fact
that
\bea 
	-\avr{\tilde g,\cL_s^*(\tilde g)}_{s,\sigma} &=&
	-\avr{g,\cL_s^*(g)}_{s,\sigma}
	+ \tr(g\sigma) \avr{\1,\cL_s^*(g)}_{s,\sigma}\nonumber\\
	&+& \tr(g\sigma) \avr{g,\cL_s^*(\1)}_{s,\sigma}
	- \tr(g\sigma)^2 \avr{\1,\cL_s^*(\1)}_{s,\sigma}
	\\
	&=&
  -\avr{g,\cL_s^*(g)}_{s,\sigma} ,
\eea
where we have used that $\cL_s^*(\1)=0$ and $s$-reversibility.
Note also that 
\bea
	\avr{\tilde g,\tilde g}_{s,\sigma} = \Var^s_\sigma (g)
\eea
and
\be 
	\left.\frac{d}{dt}\Var^s_\sigma(g_t)\right|_{t=0}=\avr{g,\cL^*(g)}_{s,\sigma} .
\ee
Hence, 
\be 
	\lambda = \inf_{g=g^\dagger} \left.-\frac{d}{dt}\log \Var^s_\sigma(g_t)\right|_{t=0} =\lambda_s,
\ee
which proves the claim.

\qed}

The same is unfortunately not true about the Log-Sobolev constant, and in general $\alpha_s\neq \alpha_{s'}$ when $s\neq s'$. For the remainder of the paper, whenever we refer to the Log-Sobolev constant without specifying $s$, we implicitly assume that we are working with the $s=\half$ case, and therefore suppress the $s$ subscript.
The Log-Sobolev constant $\alpha_\half$ is also closely related to hypercontractivity of the semi-group generated by $\cL$. It is also worth noting that $\lambda\geq\alpha$ for primitive reversible semi-groups of finite dimensional quantum systems. See Ref.~\cite{LogSobolev} for more details. 

As a direct consequence of Eqs.~(\ref{variance}) and (\ref{entropy}), we get that for any $s\in[0,1]$ and any Hermitian operator $f$ and positive operator $g$, 
\bea 
\Var^s_\sigma(f_t)&\leq& \Var^s_\sigma(f)e^{-2t\lambda_s}, \label{var:Gap}\\
\Ent^s_\sigma(g_t)&\leq&\Ent^s_\sigma(g)e^{-2t\alpha_s}.\eea
Both of these quantities lead to very simple trace norm mixing bounds for $s=\half$, as illustrated in the following theorem.

\begin{theorem}[\cite{LogSobolev}]\label{Mixingtimebound}
Let $\cL$ be a primitive $\half$-reversible Liouvillian with stationary state $\sigma$, $\chi^2$ constant $\lambda$ and Log-Sobolev constant $\alpha$.  Then, for any initial state $\rho$,  the following 
bounds hold\je{:}
\begin{enumerate}
\item{\bf $\chi^2$ bound}
\be\left\| \rho_t - \sigma \right \|_{1} \leq \sqrt{\|\sigma^{-1}\|}e^{-\lambda t}.\label{eq:tnchi}\ee

\item {\bf Log-Sobolev bound}
\be \left\| \rho_t - \sigma \right \|_{1} \leq \sqrt{2\log(\|\sigma^{-1}\|)}e^{-\alpha t}.\label{eq:tnLS}\ee
\end{enumerate} 
\end{theorem}

It should be clear from Eqs.~(\ref{eq:tnchi}) and (\ref{eq:tnLS})  that if both $\lambda$ and $\alpha$ are independent of the system size, then the Log-Sobolev bound is much stronger than the $\chi^2$ bound, as $\|\sigma^{-1}\|\geq d$, which is exponentially large in the case of a many-particle system.  We will see in the next section that this will have consequences for the nature of the correlations in the stationary state $\sigma$. Note that the assumption of reversibility, here and in the remainder of this work, could be relaxed at the cost of defining more complicated variational expressions for the $\chi^2$ and Log-Sobolev constants. 

It is natural to ask at this point how large $\|\sigma^{-1}\|$ typically is. Its value clearly depends on the specific Liouvillian, and it can certainly be engineered to be as large as desired (in principle). However, there are a number of natural situations where we can provide good estimates of $\|\sigma^{-1}\|$. We outline three below: primitive unital, thermal, and primitive free-fermionic semi-groups. For primitive unital semi-groups of a $d$-dimensional system, $\sigma=\1/d$, and hence $\|\sigma^{-1}\|=d$. For thermal semi-groups of an $N$-qubit system with Hamiltonian $H$ at inverse temperature $\beta$, the stationary state will be given by $\sigma_\beta=e^{-\beta H}/\tr{e^{-\beta H}}$. It is a straightforward calculation to see that we have the bound 
\be
d\leq\|\sigma^{-1}\|\leq de^{\beta(\| H\| -\|H^{-1}\|^{-1})}.
\ee

Given a free fermionic Liouvillian for which $\sigma$ is a full rank Gaussian fermionic stationary state,
$\|\sigma^{-1}\|$ can be directly computed from the covariance matrix $\gamma$. Clearly, the operator norm is unitarily invariant, so one can look at the normal mode
decomposition of $\sigma$. 
Transforming a vector of Majorana modes $r$ to a new vector of Majorana modes, $r\mapsto Kr$ for $K\in O(2N)$ is reflected on the level
of  covariance matrices as a transformation $\gamma\mapsto K\gamma K^T$. For a suitable orthogonal matrix $K$, one finds
\begin{equation}\label{normalmodes}
	K\gamma K^T= \bigoplus_{j=1}^n\left[
	\begin{array}{cc}
	0 & c_j\\
	- c_j & 0
	\end{array}
	\right],
\end{equation}
with $c_j\in [-1,1]$ for all $j=1,\dots, n$.
The smallest eigenvalue of the Gaussian fermionic state $\sigma $ is found to be
\begin{equation}
	\|\sigma^{-1}\|^{-1} =  \prod_{j=1}^n \frac{1-|c_j|}{2} .
\end{equation}
That is to say, one can express the smallest eigenvalue of $\sigma$ in terms of a trace function of the covariance matrix $\gamma$ 
as
\begin{equation}
	\|\sigma^{-1}\|^{-1} = \exp\left(
	\tr\left(\log \frac{1-|\gamma|}{2}\right)
	\right).
\end{equation}
Similarly, the gap $\lambda$  of a reversible free fermionic Liouvillian can be directly read off from the matrices defining the Liouvillian in terms of 
polynomials of Majorana fermions (see Refs. \cite{Prosen1,Prosen2} for more details).
It would be interesting to know how $\|\sigma^{-1}\|$ behaves for a perturbed non-primitive Liouvillian, but that question will not be addressed here.

\section{Clustering of correlations} \label{sec:CC}

A state is said to satisfy \textit{clustering of correlations} if the correlations between two distant observables decay exponentially in the distance separating them. Correlations can be quantified in many different ways in many-body systems. Here we will consider three different notions, which are useful in our setting. 
Consider the situation of a lattice system and two non-intersecting regions $A$ and $B$. 
We will write $\rho_{AB}$ to denote a state restricted to subsystems $A,B$; i.e., where the rest of the system has been traced 
out. The dimension of the Hilbert space of the physical systems belonging to $A$ and $B$ will be denoted as $d_{AB}$.
Now, let us define the following three measures of correlation between subsystems $A$ and $B$.

\begin{definition}[Correlation measures]
Let $\rho$ be a quantum state defined on the lattice $\Lambda$, and let $A,B\subset\Lambda$ be non-overlapping, then define
\begin{itemize}
\item The covariance correlation: 
\begin{equation}
	C_\rho(A:B):=\sup_{\|f\|=\|g\|=1}|\tr{[(f\otimes g)(\rho_{AB}-\rho_A\otimes\rho_B)]}|, 
\end{equation}
where $f$ is supported on region $A$, and $g$ is supported on region $B$. 
\item The trace norm correlation: 
\begin{equation}
	T_\rho(A:B):=\|\rho_{AB}-\rho_A\otimes\rho_B\|_1.
\end{equation}
\item The mutual information correlation: 
\begin{equation}	
	I_\rho(A:B):= S(\rho_{AB}\|\rho_A\otimes\rho_B), 
\end{equation}
where $S(\rho\|\sigma)=\tr[\rho(\log{\rho}-\log{\sigma})]$ is the relative entropy. 
\end{itemize}
\end{definition}

We show (in Appendix B) that these three measures of correlations are related in the following way:

\begin{proposition}[Relationship between correlation measures]\label{CCequivalences}
Let $\rho$ be a full rank state of the lattice $\Lambda$, and let $A,B\subset\Lambda$ be non-overlapping subsets. Let $d_{AB}$ be the dimension of the subsystem defined on $AB$, then the following inequalities hold,
\bea 
	\frac{1}{2d_{AB}^2}T_\rho(A:B)&\leq& C_\rho(A:B)\leq T_\rho(A:B)\label{TC},\\
	\frac{1}{2}T^2_\rho(A:B)&\leq& I_\rho(A:B)\leq \log(\|\rho_{AB}^{-1}\|)T_\rho(A:B)\label{TI}.
\eea
\end{proposition}
Proposition \ref{CCequivalences} immediately tells us that if $A,B$ are small (independent of the system size) then clustering of correlations in one of the three quantities implies clustering of correlations in the other two. However, if $A$ and $B$ are proportional to the system size, then these measures can be vastly different. 

We specifically mention the situation provided by free fermionic models, where the above relationship can be tightened, as
is shown in Appendix C. The operator norm of an observable $M = i r^T m r$ with $m=-m^T$ can then be bounded by that of
$\|m\|$. The covariance matrix $\gamma_{AB}\in \bR^{2n\times 2n}$ of $\rho_{AB}$
is a principal sub-matrix of the covariance matrix $\gamma$ of the entire quantum state $\rho$.
This covariance matrix $\gamma_{AB}$ as well as the covariance matrix $\xi_{AB}$ of the uncorrelated reductions
can be cast into the form
\begin{equation}
	\gamma_{AB} = \left[
	\begin{array}{cc}
	\gamma_A & \gamma_{C}\\
	-\gamma_C & \gamma_B 
	\end{array}
	\right], \,\,
	\xi_{AB}:= \left[
	\begin{array}{cc}
	\gamma_A & 0\\
	0& \gamma_B 
	\end{array}
	\right],
\end{equation}
respectively. Under
a transformation $r\mapsto Kr$ for $K\in O(2n)$,
the covariance matrix transforms as $\gamma_{AB}\mapsto K\gamma_{AB} K^\dagger$. Using the 
singular value decomposition, and making use of the definition of the covariance matrix as the matrix
collecting second moments of Majorana fermions, 
one hence finds the lower bound 
\begin{equation}
	C_\rho(A:B)\geq \frac{1}{2}\|\gamma_C\|,
\end{equation}
for the correlation measure $C_\rho(A:B)$ in terms of the operator norm of the off-diagonal block of the covariance matrix. 
Furthermore, the covariance correlation and the mutual information correlation can be related:

\begin{proposition}[Relationship between correlation measures for free fermionic models]\label{CCequivalencesF}
Let $\rho$ be a Gaussian fermionic quantum state defined on the lattice $\Lambda$,
and let $A,B\subset \Lambda$ with $|A|=|B|=n$ be non-overlapping subsets. Then the following inequality holds,
\begin{equation}
	I_\rho(A:B)\leq -4n \log
	\left(
 	\min \left( 
	1- \| \gamma_{AB}\|
	, 
	1-\| \xi_{AB}\|
	\right)
	\right) C_\rho(A:B).
\end{equation}

\end{proposition}


\section{Main results}\label{sec:main}

We now state and prove the main results of this work: clustering of correlation theorems for regular (and reversible) local Liouvillians with 
\begin{itemize}
\item[(i)] a $\chi^2$ constant which is independent of the system size, and 
\item[(ii)] a Log-Sobolev constant which is independent of the system size\footnote{From this point on, we will simply say $\chi^2$ (or Log-Sobolev) constant to mean a $\chi^2$ (or Log-Sobolev) constant which is independent of the system size.}. 
\end{itemize}
A very important ingredient in the proof is an open systems Lieb-Robinson bound; a tool for rigorously bounding the maximal speed of propagation of information through a lattice system. The speed of propagation implies a light cone, outside of which little
information from a local source  can be inferred
(in fact no information up to exponentially small corrections). 
Open system Lieb-Robinson bounds largely resemble
their closed system counterparts, and have been shown by a number of authors already \cite{Kliesch,LRbound1,LRbound2}. 

We shall invoke a version from Ref.~\cite{Kliesch}, in the form of a "quasi-locality of Markovian dynamics" theorem. Given a local Liouvillian $\cL=\sum_{Z\subset\Lambda}\cL_Z$,  we define the Liouvillian restricted to subsets $B$ of the lattice as $\cL_B=  \sum_{Z\subset B} \cL_Z$.

\begin{figure}
\centering
    \includegraphics[scale=0.4]{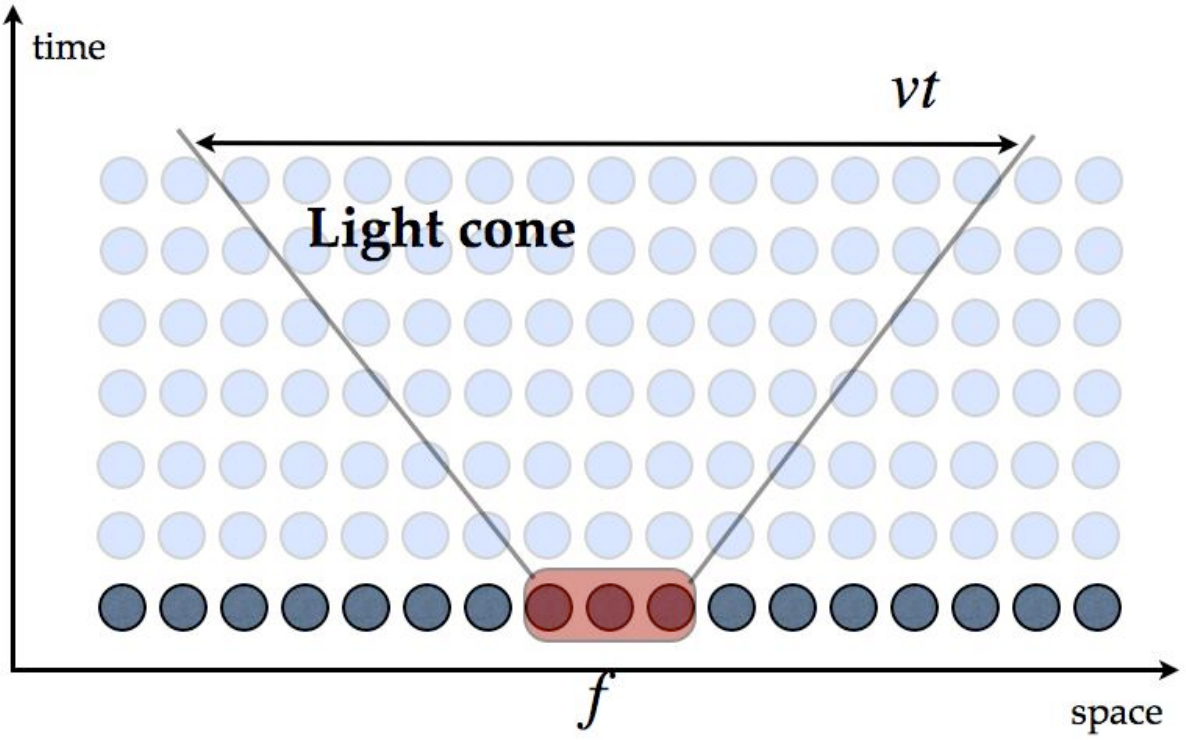}
    \caption{Depiction of the light cone of observable $f$. After a time $t$, the "information" of the observer $f$ will have propagated a distance $vt$.}
    \label{fig1a}
\end{figure}

\begin{theorem}[Open system Lieb-Robinson bound \cite{Kliesch}]\label{Kliesch} 
Let  $\cL=\sum_{Z\subset\Lambda}\cL_Z$ be a local bounded Liouvillian, and let $\cL_B$ be its restriction to the  subset $B\subset\Lambda$ of the $\cD$-dimensional cubic lattice $\Lambda$. Let $f$ be supported on $Y\subset B$, and  write the time evolved observable $f$ w.r.t. $\cL_B$ as $f^B_t$. Then for $D:=\lceil d(Y,B^c)/a\rceil$, 
\be \|f_t-f^B_t\|\leq C D^{\cD-1}\|f\| e^{vt-D},\ee for all $t>0$, where $v$ is the Lieb-Robinson velocity of $\cL$, $B^c\subset\Lambda$ is the complement of $B$, and $C>0$ is a constant which might depend upon $a$ and $\cZ$.
\end{theorem}

The theorem says that the time evolution of a local observable $f$ only depends on the terms in the Liouvillian which are in the light cone of $f$, up to an exponentially small error (see Fig. \ref{fig1a}).
The same theorem can be re-derived for free fermionic dissipative systems in the following form: Observables are taken to be
quadratic in the Majorana fermions, $F= i r^T f r$, where $f_t$ relates to the time evolved kernel of the time evolved observable $F_t$.

We now prove a corollary of this theorem which will be very useful in the following. The corollary states that for two distant observables on the lattice, it makes essentially no difference whether they evolve together or independently, as long as they are outside of each-others light cones. 

\begin{figure}
\centering
    \includegraphics[scale=0.50]{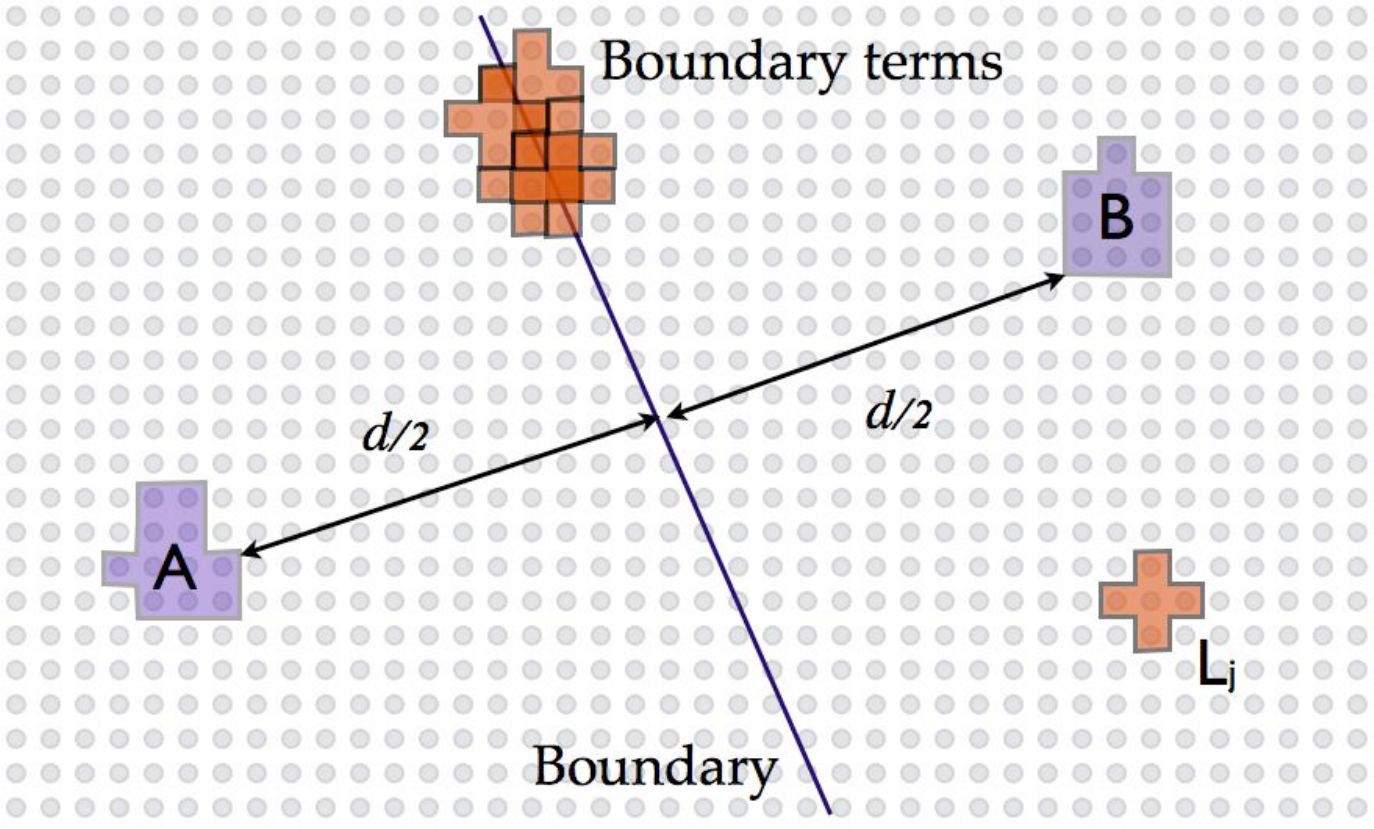}
    \caption{The subsets of the lattice $A,B\subset\Lambda$ are separated by a boundary which is halfway between the subsets $A$ and $B$. Orange crosses are meant to indicate local Lindblad operators. Lindblad operators intersecting the boundary are removed. The shapes of the regions $A$, $B$, and of the Lindblad operators are chosen for illustrative purposes only. }
    \label{fig1b}
\end{figure}

\begin{corollary}[Time evolution of spatially separated observables]\label{coro} 
Let  $\cL=\sum_{Z\subset\Lambda}\cL_Z$ be a local bounded Liouvillian. Let $A,B\subset\Lambda$ be two non-overlapping subsets of the $\cD$-dimensional cubic lattice $\Lambda$, let $f$ be be supported on $A$ and let  $g$ be supported on $B$, then 
\be \|(fg)_t-f_tg_t\|\leq C D^{\cD-1}\|f\|~\|g\| e^{vt-D/2},\ee for all $t>0$, where $v$ is the Lieb-Robinson velocity, $D:=\lceil d(A,B)/a\rceil$, and $C>0$ is a constant which might depend upon $a$ and $\cZ$.
\end{corollary}
\proof{
Let $f_t$ denote the time evolution of the observable $f$ with respect to $\cL$. Define the semi-group $\tilde{\cL}$ which is identical to $\cL$ except along a boundary $\partial_{AB}$ separating $A$ and $B$, which is chosen equidistant to the supports of $A$ and $B$, see Fig.~2. All of the local Liouvillian terms intersecting the boundary are removed $\tilde{\cL}=\cL-\sum_{Z\cap \partial_{AB}\neq0}\cL_Z$, so that $\tilde{f}_t\tilde{g}_t=\tilde{(fg)}_t$. We then get
\be 
	\|(fg)_t-f_tg_t\|\leq \|(fg)_t-\tilde{(fg)}_t\|+\|f_t g_t-\tilde{f}_t\tilde{g}_t\|\label{cor1:eq1}\je{.}
\ee
The first term on the right hand side of Eq.\ (\ref{cor1:eq1}) can be bounded by Theorem \ref{Kliesch}. For the second term, consider
\bea  
	 \|f_t g_t-\tilde{f}_t\tilde{g}_t\|&\leq& \|f_t (g_t-\tilde{g}_t)\|+\|(f_t -\tilde{f}_t)\tilde{g}_t\|\\
	&\leq& \|f\|~\|g_t-\tilde{g}_t\|+ \|f_t-\tilde{f}_t\|~\|g\|,
\eea 
where we have used that $\|f_t\|\leq\|f\|$ (shown in the appendix of Ref.~\cite{Kliesch}), and the norm Cauchy-Schwarz inequality. 
Combining all of the bounds, we get that
\be \|(fg)(t)-f(t)g(t)\| \leq C D^{\cD-1}\|f\|~ \|g\| e^{v t- D/2},\ee for some constant $C$ of order $\cO(1)$. The $D/2$ in the exponent comes from the fact that the boundary $\partial_{AB}$ lies halfway between $A$ and $B$. The possible contribution from the boundary is absorbed into the constant $C$. \qed}

Again, a free fermionic instance of this statement follows, acknowledging that the observables which are quadratic in Majorana fermions commute. It is 
still true that 
$\tilde F_t \tilde G_t = (\tilde{F G})_t$ for $F= i r^T f r$ and $G=i r^T g r$ and their time-evolved instances. 
Also, $\|F_t\|\leq \|f\|$ is still valid. Along the same lines, one hence arrives at the following
statement:

\begin{corollary}[Time evolution of spatially separated free fermionic observables]\label{coro} 
Let  $\cL=\sum_{Z\subset\Lambda}\cL_Z$ be a local bounded free-fermionic Liouvillian. Let $A,B\subset\Lambda$ be two non-overlapping subsets of the $\cD$-dimensional cubic lattice $\Lambda$, let $F= i r^T f r$ be an observable quadratic in the Majorana fermions 
supported on $A$ and similarly, let  $G=i r^T g r$ be supported on $B$, then 
\be \|(FG)_t-F_t G_t\|\leq C D^{\cD-1}\|f\|~\|g\| e^{vt-D/2},\ee for all $t>0$, where $v$ is the Lieb-Robinson velocity, $D:=\lceil d(A,B)/a\rceil$, and $C>0$ is a constant which might depend upon $a$ and $\cZ$.
\end{corollary}

\subsection{The $\chi^2$ constant}

We are now in a position to prove the first main theorem of this work.

 \begin{theorem}[$\chi^2$ clustering]
 \label{GapTheorem}
Let $A,B\subset\Lambda$ be two non-overlapping subsets of the $\cD$-dimensional cubic lattice $\Lambda$, and let $\cL=\sum_{Z\subset\Lambda}\cL_Z$ be a local bounded regular and $(0)$-reversible Liouvillian with stationary state $\sigma$, gap $\lambda$, and Lieb-Robinson velocity $v$. Then there exists a constant $c>0$ depending on $\lambda,v$ only 
such that 
\be  C_\sigma(A:B) \leq  c
	D^{\cD-1}e^{-\frac{\lambda D}{v+2\lambda}},\ee
where $D:=\lceil d(A,B)/a\rceil$.
\end{theorem}

\proof{
Let $f,g$ be Hermitian operators, with $f$ supported on $A$ and $g$ supported on $B$.  Without loss of generality, assume that $\tr[\sigma f g]\leq \tr[\sigma g f]$. Then, note that 
\be\label{thmchi:eq1}
|\Cov_\sigma(f,g)| \leq |\Cov_\sigma(f_t,g_t)| + |\Cov_\sigma(f,g)-\Cov_\sigma(f_t,g_t)|,
\ee
where $\Cov_\sigma(f,g):=\tr[f\Gamma^0_\sigma(g)]-\tr[\sigma f]\tr[\sigma g]$ defines a positive definite scalar product on hermitian operators. 
We bound the first term:
\bea  |\Cov_\sigma(f_t,g_t)| &\leq& \sqrt{\Var^0_\sigma(f_t)\Var^0_\sigma(g_t)}\\
&\leq& \sqrt{\Var^0_\sigma(f)\Var^0_\sigma(g)}e^{-2t\lambda}.
\eea
The first inequality follows from H\"olders inequality, and the second one from Eq.\ (\ref{var:Gap}). Note that we omit the $s=0$ subscript since $\lambda_0=\lambda$ is the spectral gap of $\cL$ by Lemma \ref{sLem}. Now, observe that 
\bea \label{var}
	\sqrt{\Var^0_\sigma(f)}&=& \sqrt{\tr[\sigma(f-\tr[\sigma f])^2]}\\
	&\leq& \sqrt{\|(f-\tr[\sigma f])^2\|}\\
	&\leq& \|f-\tr[\sigma f]\|\\
	&\leq& \|f\|+| \tr[\sigma f]|\leq 2 \|f\|
\eea
Therefore, we get
\be 
	|\Cov_\sigma(f_t,g_t)|\leq 4\|f\|~\|g\| e^{-2t\lambda}.
\ee
The second term in Eq.~(\ref{thmchi:eq1}) can be dealt with by invoking the theorems in the previous section. First, note that $\tr[\sigma f_t]=\tr[\sigma f]$ for any observable $f$, then
\bea |\Cov_\sigma(f,g)-\Cov_\sigma(f_t,g_t)| &=& \half(|\tr[\sigma(fg-f_t g_t)]|+ |\tr[\sigma(gf-g_t f_t)]|) \\
&=&  \half(|\tr[\sigma((fg)_t-f_t g_t)]|+|\tr[\sigma((gf)_t-g_t f_t)]|) \nonumber\\
&&\\
&\leq& \half(\|(fg)_t-f_tg_t\|+\|(gf)_t-g_tf_t\|) \\
&\leq& CD^{\cD-1}\|f\|~\|g\| e^{t v-D/2},
\eea
where in the last step we used Corollary \ref{coro}, and the assumption that the Liouvillian is regular.
We now combine both bounds and optimise for $t$. Setting $x:=CD^{\cD-1} $, we define the function $h:\bR^+\rightarrow \bR$ as
\begin{equation}
	h(t) = e^{-2\lambda t}+  x e^{vt-D/2},
\end{equation}
then the unique solution of $h'(t)=0$ is
\begin{equation}
	t_* = \frac{1}{v+2\lambda} \log\left(
	\frac{2\lambda}{xv} 
	\right) + \frac{D}{2(v+2\lambda)}.
\end{equation}	
This gives rise to the upper bound
\begin{eqnarray}
	 |\Cov_\sigma(f,g)| &\leq&  \| f\|~ \|g\|~ h(t_*) \nonumber\\
	 &=&   \| f\|~ \|g\|~\left(
	1+\frac{2\lambda}{v}
	\right)
	e^{-\frac{\lambda D}{v+2\lambda}}
	\left(
	\frac{2\lambda}{vCD^{\cD-1}}
	\right)^{\frac{-2\lambda}{v+2\lambda}}.
\end{eqnarray}
For a suitable constant $c>0$, the expression can be upper bounded by 
\bea
	 |\Cov_\sigma(f,g)| &\leq&  c \| f\|~ \|g\|~ 
	D^{(\cD-1)\left(
	\frac{2\lambda}{v+2\lambda}
	\right)}e^{-\frac{\lambda D}{v+2\lambda}}\\
	&\leq& c\| f\|~ \|g\|~ 
	D^{\cD-1}e^{-\frac{\lambda D}{v+2\lambda}}.
\eea
Taking the supremum over $\|f\|=\|g\|=1$ then completes the proof.
\qed}

\textbf{Remark}: there exist relevant Liouvillians, like Davies generators, which are $s$-reversible for all $s\in[0,1]$, but this is not true in general. It is easy to find examples of Liouvillians which are reversible for some $s\in[0,1]$ but not for another \cite{chi2}. The above theorem only requires $(0)$-reversibility, which is often the one which one would consider in practice \cite{Alicki}. 

Theorem \ref{GapTheorem} is particularly meaningful when one considers a class of Liouvillians defined on a sequence of lattices $\Lambda_N$ indexed by some natural number $N$ relating to the size of the system, whose $\chi^2$ constant 
can be lower bounded by a value independent of $N$. 

Again, a free fermionic instance is valid here. $\Cov_\sigma(F,G)$ for observables quadratic in the Majorana fermions $F=ir^T f r$ and $G=i r^T g r$ can be evaluated as before.
Eq.~(\ref{var})
is replaced by 
\begin{equation}
	\sqrt{\Var^0_\sigma(F)}\leq  \|f\|.
\end{equation}
Then the fermionic variant of the previous statement can be formulated as follows: 

 \begin{corollary}[Free fermionic $\chi^2$ clustering]
 \label{GapTheorem}
Let $A,B\subset\Lambda$ be two non-overlapping subsets of the $\cD$-dimensional cubic lattice $\Lambda$, and let $\cL=\sum_{Z\subset\Lambda}\cL_Z$ be a local bounded regular reversible free-fermionic Liouvillian with stationary state $\sigma$, $\chi^2$-constant (spectral gap) $\lambda$, and Lieb-Robinson velocity $v$. Then there exists a constant $c>0$ depending on $\lambda,v$ only 
such that 
\be  C_\sigma(A:B) \leq  c
	D^{\cD-1}e^{-\frac{\lambda D}{v+2\lambda}},\ee
where $D:=\lceil d(A,B)/a\rceil$.

\end{corollary}

Note that  the spectral gap of a free-fermionic Liouvillian can be be characterized more simply by the spectrum of the master equation for the covariance matrix. The formulation has the benefit of being exact, and only involves matrices of size $2N$ (instead of $2^N$ in the spin case). For more details see Refs.~\cite{Prosen1,Prosen2}.

\subsection{Log-Sobolev constant}

We will now consider the situation when the Log-Sobolev constant of the regular Liouvillian is independent of the size of the system, and see that we get a much stronger form of correlation decay. 
We will first need a lemma which says in colloquial terms that ``local perturbations perturb locally" \cite{Sims}. In other words, we consider local Liouvillian perturbations and look at their effect on the steady state in a region far from the perturbation. 


\begin{lemma}[Local perturbations perturb locally]
\label{lem:LSbound}
Let $A,B\subset\Lambda$ be two non-overlapping subsets of the $\cD$-dimensional cubic lattice $\Lambda$. Let $\cL=\sum_{Z\subset\Lambda}\cL_Z$ be a local primitive and $\half$-reversible Liouvillian with Log-Sobolev constant $\alpha$, and let $\cQ_A$ be a local Liouvillian perturbation, acting trivially outside of $A$. Let $\rho$ be the stationary state of $\cL$, and let $\sigma$ be the stationary state of $\cL+\cQ_A$. Then, 
\be \label{eqn:pertbound}
	\|\rho_B-\sigma_B\|_1\leq c D^{\cD-1} ({\log(\|\rho^{-1}\|)})^{1/2} e^{-\frac{\alpha D}{v+\alpha}},
\ee
where $D := \lceil d(A,B)/a\rceil$, $v$ is the Lieb-Robinson velocity, and $c>0$ is a constant which depends on $v$ and $\alpha$ only.
\end{lemma}
\proof{
First note that if $\cL$ is primitive, then $\cL+\cQ_A$ is also primitive. This follows from the fact that the Lindblad operators of $\cL$ span the entire matrix algebra (by primitivity \cite{wielandt}), so that adding more Lindblad operators cannot reduce the algebra spanned by the generator. 

We know that there exists some positive operator $0<f_B\leq\1$ such that 
\be 
	\|\rho_B-\sigma_B\|_1 = 2| \tr[(f_B \otimes \1_{B^c})(\rho-\sigma)] |,
\ee
where $B^c$ is the complement of $B$. 
Now note that for any state $\phi$ and any time $t$, we get
\bea\label{eqn:LS1lem}
 | \tr[(f_B\otimes\1_{B^c})(\rho-\sigma)] | &\leq& |\tr[(f_B\otimes\1_{B^c})(e^{t\cL}-e^{t(\cL+\cQ_A)})(\phi)]| \label{LSproofEqn}\\
&+&\half\|\tr_{B^c}[(\sigma-e^{t(\cL+\cQ_A)})(\phi)]\|_1+\half\|\tr_{B^c}[e^{t\cL}(\phi)-\rho]\|_1.\nonumber
\eea
By assumption, the Liouvillian $\cL$ satisfies a Log-Sobolev inequality, so Theorem 1 allows us to bound the last term as
\bea
	\half\|\tr_{B^c}[e^{t\cL}(\phi)-\rho]\|_1 &\leq&  \half\|e^{t\cL}(\phi)-\rho\|_1\\
	&\leq& \left(\half\log(\|\rho^{-1}\|)\right)^{1/2} e^{-t\alpha} ,
\eea
where the first inequality follows from the
monotonicity of the trace norm. For the second term in Eq.~(\ref{LSproofEqn}), note that for some $f_B:= f_B\otimes\1_{B^c}$ (from now on we suppress the $\1_{B^c}$), 
\bea
\half\|\tr_{B^c}[e^{t(\cL+\cQ_A)}(\phi)-\sigma]\|_1 &=& |\tr[f_Be^{t(\cL+\cQ_A)}(\phi-\sigma)]|\\
&\leq& |\tr[(e^{t(\cL^*+\cQ_A^*)}-e^{t\cL^*})(f_B)(\phi-\sigma)]| \nonumber\\&&+|\tr[e^{t\cL^*}(f_B)(\phi-\sigma)]|\label{pertLemeqn} .
\eea
The second term in Eq.~(\ref{pertLemeqn}) can be bounded by invoking the Log-Sobolev mixing time of $\cL$, to obtain
\bea 
|\tr[e^{t\cL^*}(f_B)(\phi-\sigma)]| &\leq& \half\|e^{t\cL}(\phi-\sigma)\|_1\\
&\leq& \half\|e^{t\cL}(\phi)-\rho)\|_1+\half\|e^{t\cL}(\sigma)-\rho)\|_1\\
&\leq&  \sqrt{2\log(\|\rho^{-1}\|)}e^{-t\alpha} .
\eea
The first term in Eq.~(\ref{pertLemeqn}) can be bounded using Lieb-Robinson bounds. Let $\tilde{\cL}_{A^c}$ 
be the restriction of $\cL$ to terms which do not intersect $A$, then
\bea
\label{eqn60}
&& |\tr[(e^{t(\cL^*+\cQ_A^*)}-e^{t\cL^*})(f_B)(\phi-\sigma)]| \leq \|(e^{t(\cL^*+\cQ_A^*)}-e^{t\cL^*})(f_B)\|~\|\phi-\sigma\|_1\nonumber\\
\\
&\leq&2\|(e^{t(\cL^*+\cQ_A^*)}-e^{t\cL^*})(f_B)\|\\
&\leq& 2\|(e^{t(\cL^*+\cQ_A^*)}-e^{t\tilde{\cL}_{A^c}^*})(f_B)\| +2\|(e^{t\tilde{\cL}_{A^c}^*}-e^{t\cL^*})(f_B)\| \\
&\leq& C D^{\cD-1}\|f_B\| e^{vt-D},
\eea
where $D:=\lceil d(A,B)/a\rceil$, and some constant $C$ from of Theorem \ref{Kliesch}.

Again, the first term in Eq.~(\ref{eqn:LS1lem}) can also be bounded using Theorem \ref{Kliesch}, in essentially the same way as above (after Eq.~(\ref{eqn60})).
\be
 	|\tr[f_B(e^{t\cL}-e^{t(\cL+\cQ_A)})(\phi)]|  \leq C \|f_B\| D^{\cD-1} e^{vt-D}, 
\ee
for some constant $C>0$.
Now, recalling that $\|f_B\|\leq 1$, we can combine the bounds in a similar way as in Theorem \ref{GapTheorem},
\begin{eqnarray}
	\|\rho_B-\sigma_B\|_1 &\leq&
	2C \|f_B\| D^{\cD-1} e^{vt-D}
	+2C D^{\cD-1}\|f_B\| e^{vt-D}\\
	&+&({8\log(\|\rho^{-1}\|)})^{1/2}e^{-t\alpha}
	+\left(2\log(\|\rho^{-1}\|)\right)^{1/2} e^{-t\alpha} 
\end{eqnarray}
Hence, there exists a constant $C_1>0$ such that 
\begin{eqnarray}
	\|\rho_B-\sigma_B\|_1\leq C_1 D^{\cD-1} e^{vt-D} + 3({2\log(\|\rho^{-1}\|)})^{1/2}e^{-t\alpha},
\end{eqnarray}
and choose the optimal $t$. Set $g:\bR^+\rightarrow \bR$ as
\begin{equation}
	g(t) = C_1 D^{\cD-1} e^{vt-D} + 3({2\log(\|\rho^{-1}\|)})^{1/2} e^{-t\alpha},
\end{equation}
then $g'(t)=0$ delivers as a unique solution
\begin{equation}
	t_*= \frac{ D+ \log\left(
	{3\alpha ({2\log(\|\rho^{-1}\|)} )^{1/2}}/{(v C D^{\cD-1})}
	\right)}{  \alpha +v}.
\end{equation}
In this way, one arrives at the bound in Eq.~(\ref{eqn:pertbound}) for some constant $c$ which depends on $v$, $\alpha$ and $\cD$. 
 \qed}
 
\textbf{Remark}: Lemma \ref{lem:LSbound} is very reminiscent of Theorem 6.9 in Ref. \cite{cubitt}, which was proved independently. Both versions of the theorem have their strengths and weaknesses. We provide a specific setting (characterized by a system-size independent Log-Sobolev constant) which allows for local stability, whereas the result in Ref. \cite{cubitt} is based on a more abstract notion that the authors call ``global rapid mixing". However, using the extra assumption of frustration freedom of the Liouvillian, Theorem 6.9 in Ref. \cite{cubitt} provides a bound which is system size independent -- a much stronger statement than ours. Whether the assumption of frustration freedom is necessary for the system size independent bound is an open question, which \je{we} consider important to resolve, but is beyond the scope of the present paper. 

We are now in a position to prove the second main theorem of this work, which gives a much stronger version of clustering of correlations when the system has a Log-Sobolev constant.

\begin{theorem}[Log-Sobolev clustering]\label{LogSobolev}
Let $A,B\subset\Lambda$ be two non-overlapping subsets of the $\cD$-dimensional cubic lattice $\Lambda$, and let $\cL=\sum_{Z\subset\Lambda}\cL_Z$ be a local bounded regular and $\half$-reversible Liouvillian with stationary state $\rho$, Log-Sobolev constant $\alpha$, and Lieb-Robinson velocity $v$, then \be 
	I_\rho(A:B)\leq c D^{\cD-1}(\log(\|\rho^{-1}\|))^{3/2} e^{-\frac{\alpha D}{2(v+\alpha)}},
\ee 
where $D:=\lceil d(A,B)/a\rceil$, and $c>0$ is a constant depending on $v$, $\cD$ and $\alpha$ only.

\end{theorem}

\proof{
To start with, as in the proof of Corollary \ref{coro}, define the semi-group $\tilde{\cL}$ which is identical to $\cL$ except along a boundary $\partial_{AB}$ separating $A$ and $B$, which is chosen equidistant to the supports of $A$ and $B$ (see Fig.~\ref{fig1b}). All of the Local Liouvillian terms intersecting the boundary are removed $\tilde{\cL}=\cL-\sum_{Z\cap \partial_{AB}\neq0}\cL_Z$, so that $\tilde{f}_t\tilde{g}_t=\tilde{(fg)}_t$. 
 By regularity, the Liouvillian $\tilde{\cL}$ is primitive. 
Now, let $\rho$ be the stationary state of $\cL$ and let $\sigma$ be the stationary state of $\tilde{\cL}$. Note that 
\bea 
	S(\rho\|\rho_A\otimes\rho_B)&=&-S(\rho)+S(\rho_A)+S(\rho_B)\\
	&\leq& -S(\rho)-\tr{[\rho_A\log{\sigma_A}]}-\tr{[\rho_A\log{\sigma_A}]}\\
	&=& S(\rho\|\sigma_A\otimes\sigma_B),
\eea where in the second line we used that $S(\rho_{A,B}\|\sigma_{A,B})\geq0$.   
Recall also that $S(\rho\|\rho_A\otimes\rho_B)=I_\rho(A:B)$ is the mutual information between subsystems $A$ and $B$. 
Now,  along the same lines as in the proof of Proposition \ref{CCequivalences}, we get
\bea
	I_\rho(A:B)&=& S(\rho_{AB}\|\rho_A\otimes\rho_B)\\
	&\leq& S(\rho_{AB}\|\sigma_A\otimes\sigma_B) \\
	&\leq& \log(\|\rho_{AB}^{-1}\|)\|\rho_{AB}-\sigma_A\otimes\sigma_B\|_1
\eea
From this point, we can apply Lemma \ref{lem:LSbound} to the trace norm to get the desired bound, by observing that the perturbation in this case is at the boundary between $A$ and $B$, which was constructed to be a distance $D/2$ away from  $A$ or $B$. \qed}

Considering Theorem \ref{LogSobolev} and Proposition \ref{CCequivalences}, we immediately see that there can in principle be a large divergence between covariance clustering and mutual information clustering. In the case of the stationary states of regular semi-groups, these two situations are characterized by the $\chi^2$ and Log-Sobolev decay constants respectively. It is important to point out, however, that for free-feemionic lattice systems, this separation does not exist. Indeed, if a free-fermionic Liouvillian has a spectral gap which is independent of the system size, then by Corollary \ref{GapTheorem} and Proposition  \ref{CCequivalencesF}, the system also satisfies mutual information clustering. This seems to be a strong indication that $\chi^2$ mixing and Log-Sobolev mixing are of the the same order for free-fermionic systems. 

\subsection{An area law for the mutual information}

We will now show an important consequence of the clustering of correlations result for the mutual information: an area law \cite{AreaReview,AreaLaw1,AreaLaw3}. 
We say that a system satisfies an \textit{area law} if for any region $A\subset\Lambda$, the mutual information between $A$ and its complement is upper bounded by a term which scales as the boundary of $A$. Such a behavior is far from obvious, as a naive bound on the mutual information will scale not as the boundary but rather
as the volume. 

\begin{theorem}[An area law for the mutual information]\label{thm:AreaLaw}
Let $\cL$ be a regular $\half$-reversible Liouvillian with stationary state $\rho$ and Log-Sobolev constant $\alpha$. Let $A\subset\Lambda$, then for any $\epsilon>0$, there exist
constants $\gamma_1,\gamma_2>0$ such that
\be I_\rho (A,A^c)\leq (\gamma_1+\gamma_2\log{\log{\|\rho^{-1}\|}}) |\partial_A|  +\epsilon, \ee
where $|\partial_A|$ is the boundary of $A$.
\end{theorem}

\proof{
The proof relies on properties of the conditional mutual information. Given a tripartition of the lattice $ABC= :\Lambda$ of mutually exclusive subsets
$A$, $B$, and $C$,
recall that the conditional mutual information of $\rho$ is given by 
\bea 
	I_\rho (A:B|C)&=& I_\rho(A:BC)-I_\rho(A:C)\\
	&=&I_\rho(AC:B)-I_\rho(B:C).
\eea
Note also that $I_\rho(A:B)\leq 2\min\{|A|,|B|\}$. That is to say, for arbitrary suitable such subsets
\bea I_\rho(A:B|C)&=&I_\rho(AC:B)-I_\rho(B:C)\\
&\leq& I_\rho(AC:B)\\
&\leq& 2\min\{|AC|,|B|\}. \eea
Now let $A\subset \Lambda$ be some connected region, let $\delta_l(x)$ be the ball of radius $l$ around the site $x$ and define 
$B_l:=\{x\in\Lambda| \delta_l(x)\cap A\neq 0, x\in A^c\}$ 
to be the ``buffer'' region of radius $l$ around $A$. Finally, denote with $C$ the remainder of the lattice. Then
\bea 
	I_\rho(A:B_lC) &=& I_\rho(A:B_l|C)+I_\rho(A:C)\\
	&\leq& 2|B_l |+ I_\rho(A:C)\\
	&\leq&2c_1l |\partial_A|+c_2 l^{\cD-1}(\log(\|\rho^{-1}\|))^{3/2} e^{-\frac{\alpha l}{2(v+\alpha)}}
\eea
for some constants $c_1,c_2>0$. Thus, if we take 
\be 
	l\geq \frac{4(v+\alpha)}{\alpha}
	\max\left\{ \log{\left(\frac{l}{\epsilon}\right)}, \log\left(\frac{c_2(\log\|\sigma^{-1}\|)^{3/2}}{\epsilon}\right)\right\},
\ee
it follows that
\bea 
	I_\rho(A:A^c)&=& I_\rho(A:B_lC)\\
	&\leq& c_1|\partial_A| \frac{8(v+\alpha)}{\alpha}\nonumber\\
	&\times&\max\left\{-\xi\left( -\frac{\epsilon \alpha}{4(v+\alpha)}\right),\log\left(\frac{c_2(\log\|\sigma^{-1}\|)^{3/2}}{\epsilon}\right)\right\}+\epsilon,
\eea
where $\xi(.)$ is the Log-product function (Lambert W-function).
Thus relabeling the constant terms $\gamma_1,\gamma_2$, we get
\be I_\rho(A:A^c)\leq (\gamma_1+\gamma_2\log{\log{\|\rho^{-1}\|}}) |\partial_A|  +\epsilon\ee
which completes the proof. \qed}

Note that it is not known whether this bound is tight or not in general. However, one would expect that in one dimension, the situation would be simpler. Indeed, as shown in Ref.~\cite{BrandaoAreaLaw} for 
closed systems in 1D, clustering of correlations in the variance is already enough to 
guarantee that the system satisfies an area law (without logarithmic corrections). This area law reminds
of the area law valid for (mixed) Gibbs states of local Hamiltonians \cite{ThermalPRL,OurCMP}, for which again no logarithmic correction is found.
Finally, we also note that Theorem \ref{thm:AreaLaw} only guarantees an area law when $\log{||\sigma^{-1}||}$ scales as a polynomial of the volume. We expect this to be the case quite generally. However, an extensive characterization of the situations when this is the case are beyond the scope of this article.


\section{Conclusion and outlook}\label{sec:concl}

\subsection{Topological order} 

An important implication of these results is that, in principle, it is possible for the stationary state of a regular Liouvillian with a $\chi^2$ constant  to have topological order, while this is not possible for regular Liouvillians which have a  Log-Sobolev constant. 

An intuitive argument for the existence of topological order in closed systems goes as follows: A pure state is topologically trivial if it can be transformed to a  classical state by a local unitary with local finite range \cite{VerstraeteHastings}, i.e., whose range grows at most as the logarithm of the system size. If this is not the case, then the state is said to have topological order.
One way of extending this notion to open systems is to say that if a given (mixed) state can be reached from  a classical state by a  TCP map of finite local range, then the state is topologically trivial. If this is not possible, then the state  has topological order. Using this definition, it is quite clear from the results in this work that the stationary state of a regular bounded local Liouvillian with a Log-Sobolev constant cannot have topological order in this sense, since the stationary state can be reached from any initial state in a time which scales as the logarithm of the system size. On the other hand, this conclusion cannot be drawn for the stationary state of a regular bounded local Liouvillian with a $\chi^2$ constant, since it can take a time linear in the system size to reach the stationary state, and there is enough time in 
principle for topological order to build up.

It is worth mentioning that a very similar notion of topological order for mixed states was introduced in Ref.\
\cite{HastingsTO}, where the criterion was instead based on the closed system analysis on a dilation space.  Indeed, any quantum dynamical semi-group can be related to a stochastic dilation, associated to a given 
Brownian motion \cite{StochDilation}. If the semi-group is local, then the stochastic dilation will be so as well, and therefore the range of the unitary dilation will be of the same order as the range of the semi-group. In this way, one can relate the mixing time of the semi-group to the definition of topological order given in Ref.~\cite{HastingsTO} in an explicit manner. Note, however, that the ancillary space of the stochastic dilation in Ref.~\cite{StochDilation} is continuous, so that further analysis is necessary for establishing a rigorous equivalence to the results in Ref.~\cite{HastingsTO}. 


We conclude this section by stating a conjecture: for a full rank state $\rho>0$, clustering of correlation in the mutual information excludes the existence of topological order, but  clustering of correlation in the covariance might still allow for it. 

\subsection{Classical simulation of stationary states and matrix-product operators}


One potential application of the findings presented here is in the classical simulation of open quantum systems, in particular for one-dimensional
models. There are a number of
approaches feasible to pursue such simulations: On the one hand, one can keep track of the open systems dynamics with a variant of the
\textit{density matrix renormalisation group} (DMRG) approach, either by evolving the mixed state in time under the Liouvillian, or to resort to a 
quantum jump approach unravelling the open systems dynamics. In the former situation, the encountered mixed states can be captured
in terms of \textit{matrix-product operators} \cite{MPO,MPO2,MPO3}. For the latter, one would formulate time evolution in terms of a classical stochastic process 
of \textit{matrix-product states}. On the other hand, one can directly turn to stationary states of primitive Liouvillians, and can formulate
a variant of DMRG to determine such states in terms of matrix-product operators. Such simulations should shed light onto 
phase diagrams in non-equilibrium. The present work suggests that if one encounters a stationary state of a primitive Liouvillian with
a Log-Sobolev constant, then such a system should be ``easy'' to simulate, in that one might conjecture
that a constant bond dimension is sufficient to approximate the stationary state with a matrix-product operator for a given error. Our work should serve as a guideline for such endeavors.

\subsection{Conclusion}

In this work, we have studied the relationship between the rate of convergence to a stationary state and the clustering of correlations for
open quantum systems described by regular local Liouvillians.
We conclude by raising the question of the how common the models which we are considering actually are? 
Indeed, the assumptions on the semi-group (boundedness, primitivity, reversibility) might seem very restrictive in view of the applications considered in 
quantum information theory. This is a valid point, as it excludes all dissipative protocols with pure or multiple fixed points. However, it is 
important to point our that the main application which we have in mind: thermal Liouvillians, are primitive and reversible. 
Also, pure stationary
states can be arbitrarily well approximated by situations captured by the theorems presented here.
Nevertheless, these results are a first indication that even for very rapidly mixing dissipative processes, there might still be interesting behavior to be seen. This work 
also gives further justification that the Log-Sobolev constant might be a better quantity to consider that the $\chi^2$ constant (spectral gap) when analyzing rapidly mixing quantum processes. It is the hope that the present results trigger further such studies.

\bigskip

\textbf{Acknowledgements}: We gratefully acknowledge fruitful discussions with T.~J.~Osborne, K.~Temme, M.~Kliesch, D.~Poulin, and M.~Friesdorf. 
We especially thank F.~G.~L.~S.~Brandao for pointing out that the mutual information clustering presented here leads to an area law. 
This work has been supported by the Alexander von Humboldt foundation,
the EU (Q-Essence, SIQS), the ERC (TAQ), the EURYI, and the BMBF (QuOReP). Similar results have very recently been obtained independently of us \cite{cubitt}. 


\pagebreak

\begin{appendix}

\section{Thermal semi-groups}

We call thermal Liouvillians, the subclass of Liouvillians which describe the dissipative dynamics resulting from the weak (or singular) coupling limit of a system coupled to a large heat bath. These Liouvillians are often called \textit{Davies generators} \cite{Davies}. See Ref.~\cite{Spohn} for a clear derivation and a discussion of when this canonical form can be assumed.   
Thermal Liouvillians can always be written as 
\be 
	\cL_\beta=\cL_0+\sum_{\omega,k}\cL_{\omega,k}.
\ee The individual terms are given by
\bea 
	\cL_0(f)&:=& i[H,f]-\half\sum_{\omega,k} \eta_k(\omega)\{S^\dag_k(\omega)S_k(\omega),f\}_+, \\
	\cL_{\omega,k} (f)&:=& \eta_k(\omega)S^\dag_k(\omega)fS_k(\omega),
\eea 
where $\omega$ are the so-called Bohr frequencies and the $k$ index  reflects the couplings to the environment. In particular, $k$ can always be chosen such that $k\leq d^2$. $\eta_k(.)$
are the Fourier coefficients of the two point correlation functions of the environment, and are bounded. The $S_k(.)$
operators can be understood as mapping eigenvectors of $H$ with energy $\omega$ to eigenvectors of $H$ with energy $E+\omega$, and hence act in the Liouvillian picture as quantum jumps which transfer energy $\omega$ from the system to the bath. 

The thermal map can be seen to have a unique (full-rank) stationary state which 
is given by $\sigma_\beta\propto e^{-\beta H}$, where $\beta$ is the inverse temperature of the heat bath. The following useful relations hold for any $k$ and $\omega$ and $s\in[0,1]$,
\bea 	\eta_k(-\omega)&=&e^{-\beta \omega}\eta_k(\omega)\label{DBDavies1},\\ 
\sigma^s_\beta S_k(\omega)&=&e^{s\beta \omega}S_k(\omega)\sigma^s_\beta \label{DBDavies2},
\eea where Eqs.~(\ref{DBDavies1}) and (\ref{DBDavies2}) are equivalent to the detailed balance condition for $\cL_\beta$.
In physical terms this means that it is as likely for the system to transfer an amount $\omega$ of energy to the environment as it is for the environment to transfer the same amount into the system. 

\section{Proof of Proposition \ref{CCequivalences}}

\proof{
The lower bound in Eq.~(\ref{TI}) is simply Pinsker's inequality, the upper bound can be obtained as:
\bea D(\rho\|\sigma) &\leq&D(\rho\|\sigma)+D(\sigma\|\rho)\\
	&=& \tr{[(\rho-\sigma)(\log{\rho}-\log{\sigma})]}\\
	&\leq& \|\rho-\sigma\|_1 \|\log{\rho}-\log{\sigma}\|\\
	&\leq&\log(\max\{\|\sigma^{-1}\|,\|\rho^{-1}\|\})\|\rho-\sigma\|_1,
\eea
which  together with fact that $\|\rho_{AB}^{-1}\|\geq\|\rho_{A}^{-1}\|~\|\rho_{B}^{-1}\|$ gives us the upper bound in Eq.\ (\ref{TI}).
The upper bound in Eq.~(\ref{TC}) can be obtained by noting that 
\begin{eqnarray} \sup_{\|f_A\|=\|f_B\|=1}|\tr{[(f_A\otimes f_B)}(\sigma_{AB}-\sigma_A\otimes\sigma_B)]|&\leq&
 \sup_{\|g\|\leq1}|\tr{[g(\sigma_{AB}-\sigma_A\otimes\sigma_B)]}|\nonumber\\
 &=&T_\sigma(A:B).
 \end{eqnarray}
And finally the lower bound is obtained by noting that we can decompose any unit norm Hermitian operator $\|f\|\leq 1$ as $f=\sum_{j,k=}^D q_{j,k} X_j\otimes Y_k$, where $q_{j,k}$ are some complex amplitudes within the unit disk, and $X_j,Y_j$ are some (Hermitian) matrix basis of the associated subspaces with $\|X_j\|_1,\|Y_j\|_1= 1$. 
Then, noting that $T_\sigma(A:B)=|\tr{[P(\sigma_{AB}-\sigma_A\otimes\sigma_B)]}|$ for some $P\leq \1$, and writing
$P=\sum_{j,k=1}^{d_{AB}}q_{j,k}X_j\otimes Y_k$ with $X_j\leq \1$ and $Y_k\leq \1$, we get
\bea \half T_\sigma(A:B)&=&|\tr{[P(\sigma_{AB}-\sigma_A\otimes\sigma_B)]}|\\
&=& |\sum_{j,k=1}^{d_{AB}}q_{j,k}\tr{[(X_j\otimes Y_k)}(\sigma_{AB}-\sigma_A\otimes\sigma_B)]|\\
&\leq& \sum_{j,k=1}^{d_{AB}}|q_{j,k}\|\tr{[(X_j\otimes Y_k)}(\sigma_{AB}-\sigma_A\otimes\sigma_B)]|\\
&\leq&d_{AB}^2 \Cov_\sigma(A:B).\eea
\qed}

\section{Proof of Proposition \ref{CCequivalencesF}}

\proof{
In terms of the covariance matrices $\gamma_{AB}$ and $\xi_{AB}$,
the mutual information is found to be a difference between two trace functions,
\begin{equation}
	I_\rho(A:B) = \tr(s(i \xi_{AB})) - \tr(s(i \gamma_{AB}) )
\end{equation}
with $s:[-1,1]\rightarrow \bR$ being defined as
\begin{equation}
	s(x) = -\frac{1+x}{2}\log_2\left(
	\frac{1+x}{2}
	\right).
\end{equation}
This expression can be easily derived: For a single mode the covariance matrix $\eta$ is a skew
symmetric real $2\times 2$ matrix with eigenvalues $\pm ic$, $c\in [-1,1]$,
while the spectrum of the corresponding
Gaussian fermionic quantum state is found to be $\{(1+c)/2, (1-c)/2\}$. The von-Neumann entropy of this single-mode state
is therefore given by 
\begin{equation}
	s(c)+s(-c)= - \frac{1+c}{2} \log_2\left(\frac{1+c}{2}\right) -  \frac{1-c}{2}  \log_2\left(\frac{1-c}{2} \right) = \tr(s(i\eta))
\end{equation}
The general result is then deduced from this by making use of a normal mode decomposition, bringing the covariance matrix $\gamma_{AB}$ 
into the form of Eq.~(\ref{normalmodes}). Note that $s:[-1,1]\rightarrow \bR$ as it is defined here is not an even function.

The mutual information will now be related to the correlation measure $C_\rho(A:B)$. 
One can make use of Weyl's perturbation theorem to see how different the spectral values of $i \gamma_{AB}$
and $i \xi_{AB}$ can possibly be. We denote the 
eigenvalues of $i \gamma_{AB}$ as $\{\lambda_1,\dots, \lambda_{2n}\}$ and the 
eigenvalues of $i \xi_{AB}$ as $\{\mu_1,\dots, \mu_{2n}\}$; they come in pairs of positive and negative values, so that 
\begin{equation}
	\lambda_j=-\lambda_{j+n},\,\,
	\mu_j=-\mu_{j+n}
	\end{equation}
for $j=1,\dots, n$.
We find, using the mean value theorem and the fact that the spectral values of covariance matrices come in pairs,
\begin{eqnarray}
	I_\rho(A:B) &=& \left| \sum_{j=1}^{2n}\left( s(\lambda_j) - s(\mu_j)\right)\right |\\
	&\leq& 
	 \sum_{j=1}^{2n}  \left|   \left(s(\lambda_j) - s(\mu_j)\right)\right |\\
	 &\leq &
	 \max \left\{|s'( \pm \lambda_j)|, |s'(\pm \mu_j)|:j=1,\dots, n\right\}   \sum_{j=1}^{2n}   |\lambda_j -\mu_j|.
\end{eqnarray}
Weyl's perturbation theorem then delivers $|\lambda_j -\mu_j|\leq \|\gamma_C\|$ for $j=1,\dots, 2n$, so that
\begin{eqnarray} 
	I_\rho(A:B)  &\leq &
	 \max ( s'( - \| \gamma_{AB}  \| ),  s'( - \| \xi_{AB}\|) )  2n \|\gamma_C\|
	 \\
	 &<&   \max \left( - 
	\log
	\left(1 - \| \gamma_{AB}  \|
	\right), -
	\log
	\left(1 -  \| \xi_{AB}  \|
	\right)
	\right)
 2n \|\gamma_C\|\\
 &=&-\log\left(
 	\min \left( 
	1- \| \gamma_{AB}\|
	, 
	1-\| \xi_{AB}\|
	\right)
	\right)
 2n \|\gamma_C\|
 .
\end{eqnarray}
Here, it has been used that 
\begin{equation}
	s'(x) = -\frac{1}{2\log(2) }\left( \log \left(
	\frac{1+x}{2}
	\right) + 1\right)< -\log(1+x)
\end{equation}
for $x\in[-1,1]$. This means that 
\begin{equation}
	I_\rho(A:B)\leq -4n \log
	\left(
 	\min \left( 
	1- \| \gamma_{AB}\|
	, 
	1-\| \xi_{AB}\|
	\right)
	\right) C_\rho(A:B).
\end{equation}
\qed}

\end{appendix}

\end{document}